\documentclass{article}%

\usepackage[intlimits]{amsmath}%
\usepackage{amsfonts}%
\usepackage{amssymb}%
\usepackage{array}%
\usepackage{braket}
\usepackage{graphicx}
\usepackage{hhline}%
\usepackage{multirow}%
\usepackage{young,youngtab}

\numberwithin{equation}{section}
\newcommand{\vc}{\Yvcentermath1}

\title{\hfill{WITS-CTP-039}\\
        \textbf{Restricted Schur Polynomials and Finite N Counting}}
\author{Storm Collins \\
		\emph{National Institute for Theoretical Physics,} \\
		\emph{Department of Physics and Centre for Theoretical Physics,} \\
		\emph{University of the Witwatersrand,} \\
		\emph{Wits, 2050, South Africa} \\
		\texttt{collins.storm@gmail.com}}

\begin{document}
\maketitle

\begin{abstract}
	Restricted Schur polynomials have been posited as orthonormal operators for the change of basis from $\mathcal{N}=4$ SYM to type IIB string theory \cite{GravI,GravII,GravIII,collins}. In this letter we briefly expound the relationship found between the restricted Schurs and the operators found by Brown, Heslop and Ramgoolam in \cite{Brown}. We then briefly examine the finite $N$ counting of the restricted Schur polynomials. 
\end{abstract}

\section{Introduction}

The AdS/CFT correspondence \cite{Maldacena, Witten, Polyakov} claims that the large $N$ limit in conjunction with the large 't~Hooft coupling limit of ${\cal N}=4$ super Yang-Mills theory is equivalent to type IIB supergravity on the AdS$_5\times$S$^5$ background. This provides a novel way of looking at the large $N$ limit of matrix models \cite{tHooft}. This is significant since solving the dynamics in the large $N$ limit of multi-matrix models is a formidable problem.

This seminal insight has already produced significant results with respect to the 1/2 BPS sector \cite{CJR,Berenstein,LLM,Balasubramanian_babel1,Balasubramanian_babel2,Brown4,GeoFromYoung}. In particular, \cite{CJR} have advocated the use of Schur polynomials (see also \cite{ramgoolam3,CorRam,Gwyn}). Their candidacy is prime since we can compute their two point function exactly and they have a simple product rule allowing us to generalize to multi-point functions.

Given the rich structure emerging from the study of the 1/2 BPS sector, it is natural to extend this to sectors with less supersymmetry. This was started by Ramgoolam and Kimura in \cite{Ramgoolam1} who studied the free field limit of the $Z,Z^\dagger$ sector of the theory. Brown, Heslop and Ramgoolam \cite{Brown} generalized this to include $M$ complex Higgs fields in a model with a global U($M$) symmetry. Using insights from excited giant graviton dynamics, an alternative description for the model with $M$ complex Higgs fields was obtained in \cite{collins}.

At least two natural questions suggest themselves. First, what is the relationship between the BHR operators and the restricted Schur polynomials? A concrete answer to this question should allow us to translate between the two bases, proving that they are equivalent. Secondly, \cite{Brown} have argued that the operators they define match the number of BPS states that can be defined in the free theory. Is this also true for the restricted Schur polynomials? The goal of this letter is to provide a concrete answer to both these questions. The change of basis is given in equation \eqref{eq:changeOfBases}. We illustrate the correctness of our result in the first nontrivial example involving 3 boxes viz. $\chi_{\text{\tiny$\vc\yng(2,1);\yng(2)\circ\yng(1)$}}$ -- rewriting it in terms of the BHR operators. Further, we count the number of restricted Schur polynomials and demonstrate agreement with Dolan \cite{Dolan}.

For completeness we end this introduction by noting that Brown, Heslop and Ramgoolam \cite{BHR2} can now do non-compact global symmetry groups and Kimura and Ramgoolam \cite{SanjayeKimura} have provided a set of Casimirs that organize the possible multi-matrix bases of operators. For a nice recent review see Ramgoolam \cite{SanjayeReview}. Finally a product rule for restricted Schur polynomials has been obtained in \cite{mike}. Together with the two-point function derived in \cite{collins} this allows us to derive higher point functions.

\section{The BHR Operator and Restricted Schur Polynomial Relationship}

We\footnote{This section was completed in collaboration with Robert de Mello Koch.} consider BHR (Brown-Heslop-Ramgoolam) operators and restricted Schur polynomials constructed from two fields labeled $X$ and $Y$, with $m$ copies of the prior and $n$ of the latter -- the extension to multi-matrices is trivial. The restricted Schur polynomials \cite{Balasubramanian2,GravI,GravII,GravIII} are given by

\begin{equation}
	\chi_{R,R_\alpha} = \frac{1}{m!n!}\sum_{\sigma \in S_{m+n}} \text{Tr}_{R_\alpha}\left(\Gamma_R(\sigma)\right) \text{Tr}(\sigma X^m Y^n), \label{eq:resSchur}
\end{equation}

where $X^m$ and $Y^n$ respectively denote $m$ copies of the field $X$ and $n$ copies of the field $Y$. $\sigma$ acts on $X^m Y^n$ by permuting the lower U($N$) indices as follows

\begin{align}
	\sigma X^m Y^n &= \sigma X^{i_1}_{j_1}X^{i_2}_{j_2}\cdots X^{i_m}_{j_m}Y^{i_{m+1}}_{j_{m+1}} Y^{i_{m+2}}_{j_{m+2}} \cdots Y^{i_{m+n}}_{j_{m+n}} \notag \\
	&= X^{i_1}_{j_{\sigma(1)}}X^{i_2}_{j_{\sigma(2)}}\cdots X^{i_m}_{j_{\sigma(m)}}Y^{i_{m+1}}_{j_{\sigma(m+1)}} Y^{i_{m+2}}_{j_{\sigma(m+2)}} \cdots Y^{i_{m+n}}_{j_{\sigma(m+n)}}.
\end{align}

$\text{Tr}_{R_\alpha}$ denotes the fact that we are only tracing over the subspace corresponding to the subduction $R_\alpha$ of the representation $R$. For more details see \cite{GravI,GravII,GravIII,collins}.

The BHR operators \cite{Brown} are given by

\begin{equation}
	\mathcal{O}^{\Lambda \mu, R'}_{\beta,\tau} = \frac{1}{(m+n)!} \sum_{\sigma \in S_{m+n}} B_{j\beta} S^{\tau, \Lambda R' R'}_{j p q} \left[ \Gamma_{R'}(\sigma)\right]_{pq} \text{Tr}(\sigma X^{m} Y^{n}),
\end{equation}

where $B_{j\beta}$ and $S^{\tau, \Lambda R' R'}_{j p q}$ denote branching and Clebsch-Gordon coefficients respectively. 

To establish a link between these two operators we proceed as follows. The trace $\text{Tr}(\sigma X^{m} Y^{n})$ can be rewritten in terms of the BHR operators as \cite{Brown} follows

\begin{gather}
	\text{Tr}(\sigma X^{m} Y^{n}) = \sum_{R'} d_{R'}
	\left[\Gamma_{R'}(\sigma) \right]_{ab} B_{c\beta} S^{\tau,\Lambda R' R'}_{c a b} \mathcal{O}^{\Lambda \mu,R'}_{\beta, \tau},
\end{gather}

Thus we can write the restricted Schur polynomials in terms of the BHR operators yielding

\begin{gather*}
	\chi_{R,R_\alpha} = \frac{1}{m!n!}\sum_{\sigma \in S_{m+n}} \text{Tr}_{R_\alpha}\left(\Gamma_R(\sigma)\right)\sum_{R'} d_{R'}
	\left[\Gamma_{R'}(\sigma) \right]_{ab} B_{c\beta} S^{\tau,\Lambda R' R'}_{c a b} \mathcal{O}^{\Lambda \mu,R'}_{\beta, \tau}.
\end{gather*}

Note that we have mixed notation here using both the restricted trace and the branching coefficients. We will rectify this shortly. The above sum can be simplified using the fundamental orthogonality relation of representations to yield

\begin{equation}
	\chi_{R,R_\alpha} = \frac{(m+n)!}{m!n!} \left[P_{R\rightarrow R_\alpha} \right]_{ba}  B_{c\beta} S^{\tau,\Lambda R R}_{c a b} \mathcal{O}^{\Lambda \mu,R}_{\beta, \tau}. \label{eq:compCG}
\end{equation}

We will convert all of the above to bra-ket notation to simplify the relationship between the BHR operators and the restricted Schurs. The projector in bra-ket notation is as follows

\begin{equation}
	\left[P_{R\rightarrow R_1\circ R_2;(\beta_1,\beta_2)}\right]_{ij} =
	\sum_{k,l}\braket{R,i|R,R_1 \circ R_2,kl,\beta_1}\braket{R,R_1\circ R_2, kl, \beta_2| R,j}, \label{eq:proj}
\end{equation}

where $\beta_1$ and $\beta_2$ label the multiplicity of $R$ in the \emph{outer} product $R_1 \circ R_2$. Similarly, the branching and Clebsh-Gordon coefficients become

\begin{gather}
	B_{c\beta} = \braket{\Lambda,c|[m]\circ[n],\beta} \label{eq:branch} \\ 
	S^{\tau,\Lambda R R}_{c a b} = \braket{R\otimes R,ab|\Lambda,\tau,c}. \label{eq:CB}
\end{gather}

By $[m]$ and $[n]$ we mean the Young diagram (and thus the corresponding representation) consisting of a single row of $m$ boxes and $n$ boxes respectively. Using equations \eqref{eq:proj}, \eqref{eq:branch} and \eqref{eq:CB} we can rewrite equation \eqref{eq:compCG} as follows

\begin{align*}
	\chi_{R,R_1\circ R_2;(\beta_1,\beta_2)} = &\frac{(m+n)!}{m!n!} \braket{R,b| R,R_1 \circ R_2,kl,\beta_1}\braket{R,R_1\circ R_2,kl,\beta_2|R,a} \\
	&\times \braket{\Lambda,c|[m]\circ[n],\beta}\braket{R\otimes R,ab|\Lambda,\tau,c}\mathcal{O}^{\Lambda\mu,R}_{\beta,\tau}.
\end{align*}

Proceeding accordingly we reduce the above to 

\begin{align}
	\chi_{R,R_1 \circ R_2;(\beta_1,\beta_2)} &= \frac{(m+n)!}{m!n!} \braket{R,R_1 \circ R_2,kl,\beta_1; R,R_1\circ R_2,kl,\beta_2 | \Lambda,\tau,c} \notag \\ &\times\braket{\Lambda,c|[m]\circ[n],\beta}\mathcal{O}^{\Lambda\mu,R}_{\beta,\tau}, \label{eq:changeOfBases}
\end{align}

which is the central result of this section. It shows that we can rewrite the restricted Schur polynomials as a linear combination of the BHR operators where 

\begin{equation}
\frac{(m+n)!}{m!n!} \braket{R,R_1 \circ R_2,kl,\beta_1; R,R_1\circ R_2,kl,\beta_2 | \Lambda,\tau,c}\braket{\Lambda,c|[m]\circ[n],\beta},
\end{equation}

acts as a Clebsch-Gordon coefficient relating the two operators. An analogy of this can be drawn. Consider a 2 dimensional harmonic oscillator. It can be quantized using either cartesian coordinates $X$ and $Y$ or the radial coordinate and angular momentum. The restricted Schur operators correspond to the first type of quantization, the BHR operators to the second. 

The restricted Schur polynomials and BHR operators corresponding to just two distinct fields (labeled $X$ and $Y$) are in fact identical. For the other operators (with no $\beta$ and $\tau$ multiplicities) it is relatively straightforward to show their equivalence if the irreducible representation is small enough.


\begin{table}[htbp]

	\begin{gather*}
		\chi_{\text{\tiny$\vc\yng(3);\yng(2)\circ\yng(1)$}} = \frac{1}{2}[\text{Tr}(X)\text{Tr}(X)\text{Tr}(Y) + \text{Tr}(XX)\text{Tr}(Y) \\
		+ 2\text{Tr}(X)\text{Tr}(XY) + 2 \text{Tr}(XXY))]
	\end{gather*}	
	
	\begin{gather*}
		\chi_{\text{\tiny$\vc\yng(2,1);\yng(2)\circ\yng(1)$}} = \frac{1}{2}[\text{Tr}(X)\text{Tr}(X)\text{Tr}(Y) + \text{Tr}(XX)\text{Tr}(Y) \\ -\text{Tr}(X)\text{Tr}(XY) - \text{Tr}(XXY)]
	\end{gather*} 

	\begin{gather*}
		\chi_{\text{\tiny$\vc\yng(2,1);\yng(1,1)\circ\yng(1)$}} = \frac{1}{2}[-\text{Tr}(X)\text{Tr}(X)\text{Tr}(Y) + \text{Tr}(XX)\text{Tr}(Y) \\ -\text{Tr}(X)\text{Tr}(XY) + \text{Tr}(XXY)]
	\end{gather*} 
	
	\begin{gather*}
		\chi_{\text{\tiny$\vc\yng(1,1,1);\yng(2)\circ\yng(1)$}} = \frac{1}{2}[-\text{Tr}(X)\text{Tr}(X)\text{Tr}(Y) + \text{Tr}(XX)\text{Tr}(Y) \\
		+ 2\text{Tr}(X)\text{Tr}(XY) - 2 \text{Tr}(XXY))]
	\end{gather*} 
	
	\caption{A list of the restricted Schur polynomials labeled by $S_3$ representations.\label{tab:restrictedSchurList}}
\end{table}

\begin{table}[htbp]

\begin{gather*}
	\mathcal{O}^{\text{\tiny$\vc\yng(3),\yng(3)$}} = \frac{1}{6}[\text{Tr}(X)\text{Tr}(X)\text{Tr}(Y) + \text{Tr}(XX)\text{Tr}(Y) \\
	+ 2\text{Tr}(X)\text{Tr}(XY) + 2\text{Tr}(XXY)]
\end{gather*}

\begin{gather*}
	\mathcal{O}^{\text{\tiny$\vc\yng(3),\yng(2,1)$}} = \frac{1}{3\sqrt{2}}[\text{Tr}(X)\text{Tr}(X)\text{Tr}(Y) - \text{Tr}(XXY)]
\end{gather*}

\begin{gather*}
	\mathcal{O}^{\text{\tiny$\vc\yng(3),\yng(1,1,1)$}} = \frac{1}{6}[\text{Tr}(X)\text{Tr}(X)\text{Tr}(Y) - \text{Tr}(XX)\text{Tr}(Y) \\
	- 2\text{Tr}(X)\text{Tr}(XY) + 2\text{Tr}(XXY)]
\end{gather*}

\begin{gather*}
	\mathcal{O}^{\text{\tiny$\vc\yng(2,1),\yng(2,1)$}} = \frac{1}{3\sqrt{2}}[\text{Tr}(XX)\text{Tr}(Y)
	- \text{Tr}(X)\text{Tr}(XY)]
\end{gather*}

	\caption{A list of the first few BHR operators. In this case there are no multiplicities so the only labels are $\Lambda$ and $R$ in $\mathcal{O}^{\Lambda,R}$. Note that the field content is the same for all the above operators i.e.\ $\mu = XXY$. \label{tab:BHRoperatorList}}
\end{table}

The easiest way to directly compute the coefficients relating the restricted Schurs to the BHR operators is through equation \eqref{eq:compCG}. We then have that

\begin{align}
	&\braket{R,R_1 \circ R_2,kl,\beta_1; R, R_1\circ R_2,kl,\beta_2 | \Lambda,\tau,c}\braket{\Lambda,c|[m]\circ[n],\beta} = \notag \\
	&\left[P_{R\rightarrow R_\alpha} \right]_{ba}  B_{c\beta} S^{\tau,\Lambda R R}_{\phantom{\tau,}c a b}.
\end{align}

Here we list the pertinent Clebsch-Gordon coefficients and representation matrices for our ensuing example computation -- these were taken from appendix D of \cite{Brown}. The Clebsch-Gordon coefficients are

\begin{gather*}
	S^{\text{\tiny$\vc\yng(3)$}RR}_{1 a b} = \frac{1}{\sqrt{d_R}}\delta_{ab}, \\
	S^{\text{\tiny$\vc\yng(2,1)$} \;\;\text{\tiny$\vc\yng(2,1)$} \;\;\text{\tiny$\vc\yng(2,1)$}}_{1 1 1} = \frac{1}{\sqrt{2}}, \\
	S^{\text{\tiny$\vc\yng(2,1)$} \;\;\text{\tiny$\vc\yng(2,1)$} \;\;\text{\tiny$\vc\yng(2,1)$}}_{1 1 1} = -\frac{1}{\sqrt{2}} \\
	S^{\text{\tiny$\vc\yng(2,1)$} \;\;\text{\tiny$\vc\yng(2,1)$} \;\;\text{\tiny$\vc\yng(2,1)$}}_{1 1 2} = S^{\text{\tiny$\vc\yng(2,1)$} \;\;\text{\tiny$\vc\yng(2,1)$} \;\;\text{\tiny$\vc\yng(2,1)$}}_{1 2 1} = 0,
\end{gather*}

and the relevant representation matrices are

\begin{gather*}
	\Gamma_{\text{\tiny$\vc\yng(2,1)$}}((1)(2)(3)) = 
	\left(\begin{array}{cc}
		1 & 0 \\
		0 & 1
				\end{array} \right) \;\;\;\;
	\Gamma_{\text{\tiny$\vc\yng(2,1)$}}((12)) = 
	\left(\begin{array}{cc}
		1 & 0 \\
		0 & -1
				\end{array} \right)
\end{gather*}

As an example let us consider the restricted Schur polynomial $\chi_{\text{\tiny$\vc\yng(2,1);\yng(2)\circ\yng(1)$}}$. This can be written in terms of BHR operators as follows

\begin{align}
	\chi_{\text{\tiny$\vc\yng(2,1);\yng(2)\circ\yng(1)$}} &= \frac{3!}{2!1!}\braket{\text{\tiny$\vc\yng(2,1),\yng(2)\circ\yng(1)$},kl;\text{\tiny$\vc\yng(2,1),\yng(2)\circ\yng(1)$},kl|\Lambda}\mathcal{O}^{\Lambda,\text{\tiny$\vc\yng(2,1)$}} \notag \\
	&= 3\braket{\text{\tiny$\vc\yng(2,1),\yng(2)\circ\yng(1)$},kl;\text{\tiny$\vc\yng(2,1),\yng(2)\circ\yng(1)$},kl|\text{\tiny$\vc\yng(3)$}}\mathcal{O}^{\text{\tiny$\vc\yng(3)$},\text{\tiny$\vc\yng(2,1)$}} \notag \\
	&+ 3\braket{\text{\tiny$\vc\yng(2,1),\yng(2)\circ\yng(1)$},kl;\text{\tiny$\vc\yng(2,1),\yng(2)\circ\yng(1)$},kl|\text{\tiny$\vc\yng(2,1)$}}\mathcal{O}^{\text{\tiny$\vc\yng(2,1)$},\text{\tiny$\vc\yng(2,1)$}}, \label{eq:continued}
\end{align}

where $\Lambda$ is determined from the direct tensor product \cite{Hamermesh} given by

\begin{gather*}
	\text{\tiny$\vc\yng(2,1)$} \otimes \text{\tiny$\vc\yng(2,1)$} = \text{\tiny$\vc\yng(3)$} \oplus \text{\tiny$\vc\yng(2,1)$} \oplus \text{\tiny$\vc\yng(1,1,1)$} 
\end{gather*}

since $\Lambda$ is a representation of U(2) only the first two Young diagrams are allowed as labels for representations. Note that the branching coefficient $\braket{\Lambda,c|[m]\circ[n],\beta}$ is missing from \eqref{eq:continued} since the multiplicity $\beta$ is 1 and thus the branching coefficient is also 1. The first coefficient is given by

\begin{align}
	\frac{3!}{2!1!}\braket{\text{\tiny$\vc\yng(2,1),\yng(2)\circ\yng(1)$},kl;\text{\tiny$\vc\yng(2,1),\yng(2)\circ\yng(1)$},kl|\text{\tiny$\vc\yng(3)$}} &= 3\left[ P_{\text{\tiny$\vc\yng(2,1)$}\rightarrow\text{\tiny$\yng(2)\circ\yng(1)$}}\right]_{ba}S^{\text{\tiny$\vc\yng(3)$}\;\;\text{\tiny$\vc\yng(2,1)$}\;\; \text{\tiny$\vc\yng(2,1)$}}_{1 a b} \notag \\
	&= 3\left(\begin{array}{cc}
				1 & 0 \\
				0 & 0
			\end{array}\right) \frac{1}{\sqrt{d_{\text{\tiny$\yng(2,1)$}}}} \delta_{ab} \notag\\
	&= \frac{3}{\sqrt{2}}.
\end{align}

The projector 

\begin{equation*}
	P_{\text{\tiny$\vc\yng(2,1)$}\rightarrow\text{\tiny$\yng(2)\circ\yng(1)$}} = \left(\begin{array}{cc} 
			1 & 0 \\
			0 & 0
			\end{array} \right),
\end{equation*}

is obtained from 

\begin{equation}
	P_{\text{\tiny$\vc\yng(2,1)$}\rightarrow\text{\tiny$\yng(2)\circ\yng(1)$}} = \frac{\left(\Gamma_{\text{\tiny$\yng(2,1)$}}((12))+1\right)}{2}.
\end{equation}

See \cite{collins} for a more exhaustive discussion of projectors. The second coefficient is obtained in a similar manner and found to be 

\begin{equation}
\frac{3!}{2!1!}\braket{\text{\tiny$\vc\yng(2,1),\yng(2)\circ\yng(1)$},kl;\text{\tiny$\vc\yng(2,1),\yng(2)\circ\yng(1)$},kl|\text{\tiny$\vc\yng(2,1)$}} = \frac{3}{\sqrt{2}}. \label{eq:exampleCalc}
\end{equation}

Using the coefficients and the polynomials in tables \ref{tab:restrictedSchurList} and \ref{tab:BHRoperatorList} we can easily verify that \eqref{eq:exampleCalc} is true.

\section{Finite N Counting of Restricted Polynomials}

In \cite{Dolan} Dolan found a formula for counting infinite $N$ BPS operators with field content $\mu = [\mu_1]\circ [\mu_2] \circ \cdots \circ [\mu_M]$ i.e.\ there are $\mu_1$ copies of field 1, $\mu_2$ copies of field 2 etc.

\begin{equation}
	N(1,2,\cdots ,M) = \sum_{R} \sum_\Lambda C(R,R,\Lambda) g(\mu;\Lambda). \label{eq:dolanCounting}
\end{equation}

Here $C(R,R,\Lambda)$ is the coefficient for $\Lambda$ in the direct product $R\otimes R$ and $g(\mu;\Lambda)$ is the Littlewood-Richardson coefficient. Note that $\Lambda$ is restricted to at most $M$ rows as a representation of $U(M)$. For finite $N$ counting one restricts the above sum over $R$ to Young diagrams with at most $N$ rows. As with the preceding section let us again restrict our discussion to at most two fields. In \cite{collins} we showed that the number of restricted Schurs at $N=\infty$ is given by

\begin{equation}
	N = \sum_{R\vdash n} \sum_{R_1 \vdash n_1} \sum_{R_2 \vdash n_2} \left(g(R_1,R_2,R)\right)^2, \label{eq:countingG2}
\end{equation}

where $g(R_1,R_2,R)$ denotes the Littlewood-Richardson coefficient of the outer tensor product $R_1 \circ R_2 = R$. $R$ is a representation of $S_n$ and $R_1$ and $R_2$ are representations of $S_{n_1}$ and $S_{n_2}$ respectively where $n_1 + n_2 = n$ since the $R_1$ and $R_2$ are subductions of the representation $R$. By this we mean (refer to equation \eqref{eq:resSchur}) there are $n_1$ copies of the field $X$ and $n_2$ copies of the field $Y$.

Using the following two identities

\begin{equation}
	g(R_1,R_2;R) = \frac{1}{n_1!n_2!} \sum_{\sigma \in S_{n_1}} \sum_{\tau \in S_{n_2}} \chi_{R_1}(\sigma) \chi_{R_2}(\tau) \chi_R(\sigma \circ \tau)
\end{equation}

and

\begin{equation}
	\sum_R \chi_R (\sigma) \chi_R(\tau) = |\text{sym}(\sigma)| \delta([\sigma]=[\tau]),
\end{equation}

we can simplify equation \eqref{eq:countingG2}. $[\sigma]$ denotes the conjugacy class of $\sigma$. By $\text{sym}(\sigma)$ we mean

\begin{equation}
	|\text{sym}(\sigma)| = i_1! 1^{i_1}i_2! 2^{i_2} \cdots i_n! 1^{i_n},
\end{equation}

where $i_j$ denotes the number of cycles of length $j$ and we have assumed $\sigma \in S_n$. The size of the conjugacy class $[\sigma]$ is given by

\begin{equation}
	|[\sigma]| = \frac{n!}{|\text{sym}(\sigma)|}.
\end{equation}

Thus equation \eqref{eq:countingG2} becomes 

\begin{align}
	N &=\frac{1}{n_1! n_2!}\sum_{\sigma_1 \in S_{n_1}} \sum_{\tau_1 \in S_{n_2}} \sum_{R\vdash n} \left(\chi_R(\sigma_1 \circ \tau_1)\right)^2 \label{eq:schurFiniteN} \\
	&= \frac{1}{n_1! n_2!} \sum_{\sigma_1 \in S_{n_1}} \sum_{\tau_1 \in S_{n_2}} |\text{sym}(\sigma_1 \circ \tau_1)|.
\end{align}

Note that the last line is only valid for $N = \infty$ counting. This last line was shown by Brown, Heslop and Ramgoolam \cite{Brown} to be related to Dolan's counting formula \eqref{eq:dolanCounting} for $N=\infty$. Thus the counting of the BHR operators and the restricted Schur polynomials do indeed coincide for $N=\infty$\footnote{We are grateful to Tom Brown for an email correspondence in which this argument first appeared.}. 

For finite $N$ counting of the restricted Schur polynomials we just restrict the sum over $R$ in equation \eqref{eq:schurFiniteN} to Young diagrams with at most $N$ rows in them. For an explicit link between the finite $N$ counting of restricted Schur polynomials and the finite $N$ counting of BPS operators consider the following.

We\footnote{The finite $N$ counting arguments in the remainder of this section first appeared in \cite{DMKunpublished}.} will study the finite $N$ counting of the restricted Schur operators for the case that we have $n-1$ $X$ fields and just 1 $Y$ field. In this case we subduce $R$ by removing only 1 box. The connection between the finite $N$ counting of the two different operators becomes apparent when we consider the following bijective mapping between subductions and the multi-trace gauge invariant operators. We can translate the individual traces over the $n-1$ $X$ fields and the 1 $Y$ field in the operator into columns of a Young diagram. A few examples will make this clear. Consider

\begin{equation}
	\text{Tr}(XXY)\text{Tr}(XX) = \young(\hfill\hfill,\hfill\hfill,\bullet),
\end{equation}

here $\text{Tr}(XXY)$ corresponds to the first column of the Young diagram and the box labeled with the bullet corresponds to the $Y$ field. Here are a few more examples

\begin{gather}
	\text{Tr}(XX)\text{Tr}(XY) = \young(\hfill\hfill,\hfill\bullet) \;\;\;\; \text{Tr}(XXY) = \young(\hfill,\hfill,\bullet) \\
	\text{Tr}(XXX)\text{Tr}(XX)\text{Tr}(Y) = \young(\hfill\hfill\bullet,\hfill\hfill,\hfill)
\end{gather}

By equation (3.5) of \cite{Dolan} we see that this yields a valid link between the finite $N$ counting of BPS operators and the finite $N$ counting of restricted Schur polynomials. 

After this warm up, we show that the number of finite $N$ restricted Schur polynomials is in agreement with the finite $N$ partition function of Dolan \cite{Dolan}. This partition function is given by

\begin{align}
	Z_{U(N)}(t) = \frac{1}{(2\pi i)^N N!} \oint \prod_{i=1}^N \frac{dz_i}{z_i}\Delta(z) \Delta(z^{-1}) \prod_{j=1}^k \prod_{r,s=1}^N \frac{1}{1- t_j z_r z_s^{-1}}. \label{eq:finiteNPartitionFunction}
\end{align}

To evaluate this we use the Cauchy-Littlewood formula which is given by

\begin{equation}
	\prod_{i=1}^L \prod_{j=1}^M \frac{1}{1- x_i y_i} = \sum_{\substack{\lambda \\ l(\lambda)\leq \min(L,M)}} \chi_\lambda(x_1, \cdots, x_L) \chi_\lambda(y_1, \cdots, y_M),
\end{equation}

where $\chi_\lambda$ denotes a Schur polynomial. Thus equation \eqref{eq:finiteNPartitionFunction} becomes

\begin{align}
	Z_{U(N)}(t) = \frac{1}{(2\pi i)^N N!} \oint \prod_{i=1}^N \frac{dz_i}{z_i}\Delta(z) \Delta(z^{-1}) \prod_{j=1}^k \sum_{\substack{\lambda_j \\ l(\lambda_j)\leq N}} \chi_{\lambda_j}(t_jz) \chi_{\lambda_j} (z^{-1}) \label{eq:finiteNPart2}. 
\end{align}

The Schur polynomial can be written as 

\begin{equation}
	\chi_{\lambda_j}(t_j z) = t_j^{|\lambda_j|} \chi_{\lambda_j}(z).
\end{equation}

Utilising this in equation \eqref{eq:finiteNPart2} and swapping the product and the sum we have
 
\begin{align}
	Z_{U(N)}(t) &= \frac{1}{(2\pi i)^N N!} \sum_{\substack{\lambda_1, \lambda_2, \cdots, \lambda_k \\ l(\lambda_i) \leq N}} (t_1)^{|\lambda_1|}(t_2)^{|\lambda_2|} \cdots (t_k)^{|\lambda_k|} \notag \\ 
	&\times \oint \prod_{i=1}^N \frac{dz_i}{z_i} \Delta(z) \Delta(z^{-1}) \chi_{\lambda_1}(z)\chi_{\lambda_1}(z^{-1}) \chi_{\lambda_2}(z)\chi_{\lambda_2}(z^{-1}) \notag \\
	&\cdots \chi_{\lambda_k}(z)\chi_{\lambda_k}(z^{-1}).
\end{align}

Using the product rule for Schur polynomials we get

\begin{align}
	Z_{U(N)}(t) &= \frac{1}{(2\pi i)^N N!} \sum_{\substack{\lambda_1, \cdots, \lambda_{k+2}\\ l(\lambda_i) \leq N}} (t_1)^{|\lambda_1|}(t_2)^{|\lambda_2|} \cdots (t_k)^{|\lambda_k|} \notag \\ 
	&\times g_{\lambda_1 \lambda_2 \cdots \lambda_k \lambda_{k+1}} g_{\lambda_1 \lambda_2 \cdots \lambda_k \lambda_{k+2}} \notag \\
	&\times \oint \prod_{i=1}^N \frac{dz_i}{z_i} \Delta(z) \Delta(z^{-1}) \chi_{\lambda_{k+1}}(z)\chi_{\lambda_{k+2}}(z^{-1}).
\end{align}	

In the above $g_{\lambda_1 \lambda_2 \cdots \lambda_k \lambda_{k+1}}$ counts how many times $\lambda_{k+1}$ appears in the product $\lambda_1 \otimes \lambda_2 \otimes \cdots \otimes \lambda_k$. These can be expressed in terms of Littlewood-Richardson coefficients. For example if $k=4$ 

\begin{equation}
	g_{\lambda_1\lambda_2\lambda_3\lambda_4} = \sum_\lambda g_{\lambda_1\lambda_2\lambda}g_{\lambda\lambda_3\lambda_4},
\end{equation}

where $g_{\lambda_1\lambda_2\lambda}$ is the usual Littlewood-Richardson coefficient.Using the inner product employed by Dolan \cite{Dolan} we have

\begin{align}	
	Z_{U(N)}(t) &= \sum_{\substack{\lambda_1, \cdots, \lambda_{k+2} \\ l(\lambda_i) \leq N}} (t_1)^{|\lambda_1|} \cdots (t_k)^{|\lambda_k|} g_{\lambda_1 \lambda_2 \cdots \lambda_k \lambda_{k+1}} \notag \\ 
	&\times g_{\lambda_1 \lambda_2 \cdots \lambda_k \lambda_{k+2}} \braket{\chi_{\lambda_{k+1}}(z),\chi_{\lambda_{k+2}}(z^{-1})} \notag \\
	&= \sum_{\substack{\lambda_1, \cdots, \lambda_{k+2} \\ l(\lambda_i) \leq N}} (t_1)^{|\lambda_1|} \cdots (t_k)^{|\lambda_k|} g_{\lambda_1 \lambda_2 \cdots \lambda_k \lambda_{k+1}} \notag \\ 
	&\times g_{\lambda_1 \lambda_2 \cdots \lambda_k \lambda_{k+2}} \delta_{\lambda_{k+1},\lambda_{k+2}} \notag \\
	&= \sum_{\substack{\lambda_1, \cdots, \lambda_{k+1} \\ l(\lambda_i) \leq N}} (t_1)^{|\lambda_1|} \cdots (t_k)^{|\lambda_k|} \left(g_{\lambda_1 \lambda_2 \cdots \lambda_k \lambda_{k+1}}\right)^2.
\end{align}

The last line is exactly the counting function for the number of finite $N$ restricted Schur polynomials. Thus the number of finite $N$ restricted Schur polynomials is in agreement with the finite $N$ partition function.

As $N \rightarrow \infty$ we get

\begin{equation}
	\sum_{\lambda_1 \cdots \lambda_{k+1}} g_{\lambda_1 \cdots \lambda_{k+1}} t_1^{|\lambda_1|} \cdots t_1^{|\lambda_k|} = \prod_{k=1}^\infty \frac{1}{1- \left(t_1^k +t_2^k + \cdots + t_m^k\right)}
\end{equation}

This generalizes theorem 4.1 of Willenbring \cite{Willenbring}.

\section*{Acknowledgments} 

I would like to thank my MSc supervisor Robert de Mello Koch, who without his great deal of help, effort, suggestions and ongoing support this paper would not have been possible. I would like to thank Tom Brown for useful correspondence. I would also like to thank Michael Stephanou and Norman Ives for reviewing this letter. 

This work is based upon research supported by the South African Research Chairs Initiative of the Department of Science and Technology and National Research Foundation. Any opinion, findings and conclusions or recommendations expressed in this material are those of the authors and therefore the NRF and DST do not accept any liability with regard thereto.

\bibliographystyle{JHEP}
\bibliography{references}	
\end{document}